\documentstyle[preprint,aps]{revtex}
\begin{document}
\draft
\title{Temperature Dependence of Pair Correlations in Nuclei in the 
Iron-Region}

\author{K. Langanke$^1$, D. J. Dean$^{1,2}$, P. B. Radha$^1$, 
and S. E. Koonin$^1$}
\address{$^1$W. K. Kellogg Radiation Laboratory, 106-38, California
Institute of Technology\\ Pasadena, California 91125 USA \\
$^2$ Physics Division, Oak Ridge National Laboratory, P.O. Box 2008\\
Oak Ridge, Tennessee 37831 USA }
\date{\today}
\maketitle

\tightenlines

\begin{abstract}
We use the shell model Monte Carlo approach 
to study thermal properties and pair correlations in 
$^{54,56,58}$Fe and in $^{56}$Cr.
The calculations are performed   
with the modified Kuo-Brown interaction in the complete $1p0f$  
model space. We find generally that the proton-proton
and neutron-neutron $J=0$ pairing correlations, which dominate the ground state
properties of even-even nuclei, vanish at temperatures around 1 MeV.
This pairing phase transition is
accompanied by  
a rapid increase in the moment of inertia and a partial unquenching of 
the M1 strength. We find that the M1 strength totally unquenches at higher
temperatures, related to the vanishing of isoscalar proton-neutron
correlations, 
which persist to higher temperatures than the pairing between
like nucleons. The Gamow-Teller strength is also
correlated to the isoscalar proton-neutron pairing and hence also unquenches
at a temperature larger than that of the pairing phase transition. 
\end{abstract}

\narrowtext

\section{Motivation}

It has long being recognized that pairing correlations play an essential
role in low-energy nuclear physics. In fact, mean-field theories \cite{Ring}
like Hartree-Fock+BCS (which adds pairing as an additional
degree of freedom to the mean-field solution) or the more consistent
Hartree-Fock-Bogoliubov theory have proven to be quite successful
in the general description of ground states of even-even nuclei. 
Although in many cases
pairing among like nucleons is sufficiently well approximated  
by Cooper pairs
(in which a nucleon is coupled to its time-reversed partner) 
phenomenology indicates that d-wave pairs of like nucleons play an
important role in deformed heavier nuclei, such as 
those of the rare-earth region.
This observation prompted the development of the Interacting Boson Model,
which is remarkably successful in describing low-lying nuclear
spectra \cite{Arima}.

The properties of nuclei at finite temperature have attracted recent
experimental \cite{exp} and theoretical \cite{theory,egido,goodman} interest.
Although it is obvious that the influence of pairing decreases with
temperature and must vanish 
in the high-temperature limit, the mean-field
description of nuclei at finite temperature
is inadequate due to the neglect of important
quantum and statistical fluctuations \cite{theory}. 
Recently developed shell model Monte Carlo (SMMC)
methods \cite{Lang} do not have this shortcoming and also
allow the consideration of model spaces large enough to account for the
relevant nucleon-nucleon correlations at low and moderate temperatures
($T \leq 1.4$ MeV \cite{temp}).

A first study considered
the thermal properties of $^{54}$Fe
within the SMMC approach \cite{temp}. 
An important result was that 
the like-pair
correlations, described by the BCS-like proton and neutron monopole
pairs, vanish  
in a
small temperature interval around $T=1$ MeV. 
This pairing phase transition is accompanied by
a rather sharp rise of the moment of inertia and a partial unquenching
of the $B(M1)$ strength. However, we also found that the total Gamow-Teller
strength $B(GT_+)$ remains roughly constant 
through
the phase transition and does not unquench. It was  conjectured
\cite{temp} that the isoscalar proton-neutron correlations, which are known
to be the main source for the quenching of the $B(GT_+)$ strength 
\cite{Engel,GT}, persist to higher temperatures than the like-pair correlations.

The purpose of this paper is to demonstrate the
validity of that conjecture.
To that end we define improved measures for the strength of the
various pair correlations and study their temperature dependence,
as well as their correlation with such nuclear observables as the moment
of inertia, and the total $B(M1)$ and $B(GT_+)$ strengths. Detailed 
SMMC calculations
are presented for the nuclei
$^{54,56,58}$Fe and $^{56}$Cr.

Our paper is organized as follows. After a brief discussion of the SMMC
method in Section 2, we introduce our definition of the pairing 
correlation strengths. In Section 3 we present the results of our 
study. We begin by discussing the temperature dependence of the moment
of inertia, the $B(M1)$ and $B(GT_+)$ strengths, and the monopole
pairing fields for all four nuclei, demonstrating that the results found in
Ref. \cite{temp} appear to be a general feature of even-even nuclei with
$A \approx 56$. In Section 3.B we study the temperature dependence
of the various pair correlations in the four nuclei, paying particular
attention to the phase transition related to the breaking of monopole pairs
as found in \cite{temp} and the presence of
proton-neutron correlations
at higher temperatures. Finally we relate in Section 3.C 
the calculated temperature dependences of the moment of inertia and the
$B(M1)$ and $B(GT_+)$ strengths to the temperature dependences of
selected pairing correlations.

\section{Theoretical background}

We exploit recently developed Monte Carlo techniques  
\cite{Lang,Johnson} to calculate the thermal properties of 
$^{54,56,58}$Fe and $^{56}$Cr in a complete  
$0\hbar\omega$ model space with a realistic interaction. The methods  
we use describe the nucleus by a canonical ensemble at temperature  
$T=\beta^{-1}$ and employ a Hubbard-Stratonovich linearization of the  
imaginary-time many-body propagator, $e^{-\beta H}$, to express  
observables as path integrals of one-body propagators in fluctuating  
auxiliary fields \cite{Lang,Alhassid}. Since Monte Carlo techniques  
avoid an explicit enumeration of the many-body states, they can be  
used in model spaces far larger than those accessible to conventional  
methods. The Monte Carlo results are in principle exact and are in  
practice subject only to controllable sampling and discretization  
errors. 
To circumvent the notorious ``sign problem'' encountered in the Monte
Carlo shell model calculations with realistic interactions \cite{Alhassid},
we adopted the procedure suggested in Ref. \cite{temp}  
which is based on an  extrapolation from a family of
Hamiltonians that is free of the sign problem
to the physical Hamiltonian. 

Our calculations include the complete set of  
$1p_{3/2,1/2}0f_{7/2,5/2}$ states interacting through the realistic  
Kuo-Brown Hamiltonian \cite{Kuo} as modified in \cite{Zuker}.
Some $5\times10^9$  
configurations of the valence neutrons and protons moving  
in these 20 orbitals are involved in the canonical ensemble.
As in Ref. \cite{temp}, 
the results presented below have been obtained in MC shell model  
studies with a time step of $\Delta\beta=1/32~{\rm  
MeV}^{-1}$ using about 5000 independent Monte Carlo samples at six values  
of the coupling constant $g$ spaced between $-1$ and 0 and the value
$\chi=4$ in the decomposition of the Hamiltonian.  
A linear
extrapolation to the physical case ($g=1$) has been adopted
for the observables discussed below.

The results presented in this paper correspond to various nuclear observables
at selected temperatures: the total $B(M1)$ and Gamow-Teller strengths and
the moment of inertia $I$, given by $I=3 \langle J^2 \rangle/T$. The total
$B(M1)$ strength is defined as $B(M1)=\langle {\vec \mu}^2 \rangle$, where
the magnetic moment $\vec \mu$ is given by ${\vec \mu} = \sum_i \mu_N
\left\{ g_l {\vec l} + g_s {\vec s} \right\}$, $\mu_N$ is the nuclear
magneton and $g_l,g_s$ are the free gyromagnetic ratios for angular
momentum and spin, respectively ($g_l=1, g_s=5.586$ for protons and
$g_l=0, g_s=-3.826$ for neutrons). The total Gamow-Teller strength is given
by $B(GT_+)=\langle({\vec \sigma} \tau_+)^2 \rangle$. We also explore the 
isovector monopole pairing, as described in Section 3.A.

The main focus of this paper is on pairing correlations and their relation
to nuclear observables. In our complete $0 \hbar \omega$ $fp$-shell model
space, a pair of protons or neutrons
with angular momentum quantum numbers $(JM)$
is generated by ($a=\pi$ for protons and $a=\nu$ for neutrons)
\begin{equation}
A_{JM}^\dagger (j_a,j_b) = 
\frac{1}{\sqrt{1+\delta_{j_a j_b}}}
\left[ a_{j_a}^\dagger a_{j_b}^\dagger \right]_{(JM)}
\end{equation}
where $\pi_j^\dagger$ ($\nu_j^\dagger$)
creates a proton (neutron) in an orbital with total spin $j$.
Proton-neutron pairs can have either isospin $T=1$ or $T=0$. The respective
definitions for the pair operators are
\begin{equation}
A_{JM}^\dagger (j_a,j_b) = 
\frac{1}{2 \sqrt{1+\delta_{j_a j_b}}}
\left[ 
\nu_{j_a}^\dagger \pi_{j_b}^\dagger +
\pi_{j_a}^\dagger \nu_{j_b}^\dagger 
\right]_{(JM)}
\end{equation}
for the isovector proton-neutron pairs and
\begin{equation}
A_{JM}^\dagger (j_a,j_b) = 
\frac{1}{2 \sqrt{1+\delta_{j_a j_b}}}
\left[ 
\nu_{j_a}^\dagger \pi_{j_b}^\dagger -
\pi_{j_a}^\dagger \nu_{j_b}^\dagger 
\right]_{(JM)}
\end{equation}
for the isoscalar pair.

A general description of a pair, $\Delta_{JM}^\dagger$,
is then given by a superposition of the pair operators:
\begin{equation}
\Delta_{JM}^\dagger = \sum_{j_a \geq j_b} \beta_{JM} (j_a,j_b) A_{JM}^\dagger
(j_a,j_b).
\end{equation}
In mean-field theories like Hartree-Fock+BCS or HFB, the weights $\beta$
can be related to the two-body potential matrix elements by the variational
minimization of the free energy in the mean-field model space. However,
such a relation is not to our disposal in the complete shell model and
thus the weights are apriori undetermined. To overcome this arbitrariness
of the definition (4) we will define the pairing strength
as follows. With (1-3) we build a generalized pair matrix
\begin{equation}
M_{\alpha \alpha'}^J = \sum_M \langle A_{JM}^\dagger (j_a,j_b)
A_{JM} (j_c,j_d) \rangle
\end{equation}
which corresponds to the calculation of the canonical ensemble average
of two-body operators like $\pi_1^\dagger \pi_2^\dagger \pi_3 \pi_4$.
The index $\alpha$ distinguishes the various possible $(j_a,j_b)$ 
combinations (with $j_a \geq j_b$).
For example, the square matrix $M$ has 
dimension $N_J=4$ for $J=0$, $N_J=7$ for $J=1$, $N_J=8$ for $J=2,3$.
In the second step, the matrix $M^J$ is diagonalized ($\beta = 1, ..., N_J$),
labelling the eigenvectors and eigenvalues in decreasing order
\begin{equation}
M^J \chi_{\beta} = \lambda_{\beta}^J \chi_{\beta}  . 
\end{equation}

\noindent
The presence of a pair condensate in a correlated ground state will be
signaled by the largest eigenvalue for a given $J$, $\lambda_{1}^J$,
being much greater than any of the others.

To discuss the temperature dependence of pairing correlations
it is convenient to introduce an overall measure for the pairing strength.
We define the pairing strength for pairs with spin $J$ as the sum
of the eigenvalues of the matrix $M^J$,
\begin{equation}
P^J = \sum_{\beta} \lambda_{\beta}^J = \sum_{\alpha} M^J_{\alpha \alpha}.
\end{equation}
We note that for a model space spanned by a single $j$-shell one finds
a simple sum rule for proton and neutron pairs
\begin{equation}
\sum_J P^J = \frac{1}{2} N_v (N_v-1),
\end{equation}
where $N_v$ is the number of valence protons or neutrons in the shell.
This sum rule holds approximately at low temperatures to the protons
of the nuclei studied in this paper, as the valence protons then occupy
mainly the $f_{7/2}$ orbital. The sum rule is not valid at higher
temperatures as the valence particles are then spread out over several
orbitals.

With our definition (7) the pairing strength is non-negative and indeed
positive at the 
mean-field level. As
we will focus on the pairing correlations beyond the
mean field in this paper, we also calculated the mean-field pairing
strength, $P_{\rm mf}^J$ by the same procedure as outlined above,
however, replacing the expectation values of the two-body matrix
elements in the definition of the generalized pair matrix $M^J$ by
\begin{equation}
\langle a_1^\dagger a_2^\dagger a_3 a_4 \rangle \rightarrow
n_1 n_2 \left( \delta_{13} \delta_{24} - \delta_{23} \delta_{14} \right)
\end{equation}
where $n_k = \langle a_k^\dagger a_k \rangle$ is the occupation number
of the orbital $k$. 

We then finally define the pairing correlation in the nucleus as the
pairing strength beyond the mean field, 
\begin{equation}
P^J_{\rm corr} = P^J - P^J_{\rm mf}.
\end{equation} 

\section{Results}

\subsection{Signals for the phase transition}

Ref. \cite{temp} reported on SMMC calculations of the thermal properties
of $^{54}$Fe within the complete $pf$-shell. One
quantity studied was the thermal response of BCS-like monopole-pair correlations
among protons and neutrons, defined as ($a=\pi,\nu$ for protons and neutrons,
respectively)
\begin{equation}
\Delta_{BCS}^\dagger = \sum_{m>0} a_m^\dagger a_{\bar m}^\dagger .
\end{equation}
It was observed that these BCS-like pairs 
break at temperatures near 1 MeV. Of the observables studied in Ref. 
\cite{temp} three exhibit a particularly interesting behavior
with increasing temperature near
the phase transition: a) the moment of inertia $I$ rises sharply; b) the
M1 strength $B(M1)$ shows a sharp rise, but unquenches only partially; and c)
the Gamow-Teller strength $B(GT_+$) 
remains roughly constant and strongly quenched.

We will now show that the breaking of BCS-like pairs at temperatures
near 1 MeV and the related behavior of the three observables (moment of
inertia and M1 and Gamow-Teller strengths) is not a particular result for
$^{54}$Fe, but is a more general feature of even-even nuclei in the
$A=56$ mass region. To this end, we present the results
of SMMC calculations for $^{56}$Cr and $^{54,56,58}$Fe
at selected temperatures. Fig. 1 shows the temperature dependence
of $I$, $B(M1)$ and $B(GT_+$)
for all four nuclei. Additionally, the expectation values for BCS-like
pairing fields, $\langle \Delta_{\rm BCS}^\dagger \Delta_{\rm BCS} \rangle$,
for both protons and neutrons are plotted as functions of temperature.
It is obvious from Fig. 1
that the like-pair correlations disappear at around 1 MeV for all four nuclei.
This phase transition is accompanied by the same behavior of $I$,
$B(M1)$ and $B(GT_+)$ 
as discussed in Ref. \cite{temp} and summarized in points a)-c) above.
We note that Ref. \cite{temp} employed a different residual interaction
(the Brown-Richter interaction \cite{Richter}) 
than used here; the temperature dependence
of the quantities shown in Fig. 1 are apparently largely independent of the
interaction. 

While Ref. \cite{temp} presented results only up to $T=2$ MeV, we have
continued the calculations in this paper to higher temperatures. Although
the quantitative results are expected to be affected by our finite model
space at temperatures larger than about 1.4 MeV \cite{temp}, 
the qualitative features of the observables are likely still
meaningful. The moment of inertia shows a $T^{-1}$-decrease with temperature
at $T > 2 $ MeV, as is expected from the definition
$I=3 \langle J^2 \rangle/T$; our finite model space
requires $\langle J^2 \rangle$ to approach a constant in the
high-temperature limit. The $B(M1)$ strength unquenches in two steps.
Following its partial unquenching at the pairing phase transition, it
remains roughly constant up to $T =2.6$ MeV, where it 
finally starts to fully unquench. Although not shown in Fig. 1,
we have checked that in the high-temperature limit the $B(M1)$ strengths
approach the appropriate values. Being roughly
constant at lower temperatures, $B(GT_+)$ unquenches at
$T>2.6$ MeV. We observe that this happens simultaneously with the second
step of the unquenching of $B(M1)$ and, as we will see below,
the two phenomena have a common
origin. Again we have checked that $B(GT_+)$ 
approaches the appropriate values in the high-temperature limit.

\subsection{Pair correlations}

We have studied the pair correlations in the four nuclei for the various
isovector and isoscalar pairs up to $J=4$. Detailed calculations have been
performed for selected temperatures between $T=0.5$ MeV and 8 MeV.
As has been shown in Refs. \cite{Alhassid,fpshell}, the SMMC studies at $T=0.5$ MeV
approximate the ground state properties for even-even nuclei well.
We have checked that in the high-temperature limit, which in our calculations
corresponds to $T=1/(\Delta \beta) = 32$ MeV, the pairing correlations
approach the appropriate mean-field values.

Although it is the pairing correlation
$P_{\rm corr}^J$ (i.e.,  
pairing strength relative to
the mean-field) that matters for the behavior of the physical observables,
we will at first discuss some features of the pair matrix $M$ and its
spectrum.
In Fig. 2 we show the pair matrix eigenvalues $\lambda^J_{\beta}$
for the three isovector $J=0^+$  
and the isoscalar $1^+$ pairing channels as calculated for the iron 
isotopes $^{54-58}$Fe. We compare the SMMC results with those derived
on the mean-field level, as discussed in section 2.
Additionally, Fig. 2 shows the diagonal matrix elements of the
pair matrix $M_{\alpha,\alpha}$, where we use the notation
$\alpha=1,..,4$ for $J=0^+$ pairs in the $f_{7/2}$, $p_{3/2}$, $f_{5/2}$
and $p_{1/2}$ orbitals. For the isoscalar pairs only the three largest
diagonal matrix elements of $M$ are shown corresponding to pairs
in $(f_{7/2})^2$ ($\alpha=1$), $(p_{3/2})^2$ ($\alpha=2$) and
$(f_{7/2} f_{5/2})$ $(\alpha=3$) proton-neutron configurations.
As expected, the protons occupy mainly
$f_{7/2}$ orbitals in these nuclei. Correspondingly, the 
$\langle A^\dagger A \rangle$
expectation value, $M_{11}$, is large for this orbital;
the other diagonal elements of $M$ are small.
For neutrons, the pair matrix is also largest for the $f_{7/2}$ orbital.
The excess neutrons in $^{56,58}$Fe occupy the $p_{3/2}$ orbital,
signalled by a strong increase of the corresponding pair matrix element
$M_{22}$ in comparison to its value for $^{54}$Fe. Upon closer inspection
we find that the proton pair matrix elements are not constant within the
isotope chain. This behavior is mainly caused by the isoscalar proton-neutron
pairing. The dominating role is played by the isoscalar $1^+$ channel, which
couples protons and neutrons in the same orbitals and in spin-orbit partners.
We thus find for $^{54,56}$Fe that the proton pair matrix in the
$f_{5/2}$ orbital, $M_{33}$, is larger than in the $p_{3/2}$ orbital,
although the latter is favored in energy. For $^{58}$Fe, this ordering
is inverted, caused by 
the increasing number of neutrons in the $p_{3/2}$ orbital which in turn
also increase the proton pairing in this orbital.

After diagonalization the proton pairing strength is essentially found
in one large eigenvalue. Furthermore we observe that this eigenvalue is
significantly larger than the largest eigenvalue on the mean-field level
supporting the existence of a proton pair condensate in the
ground state of these nuclei. 
For the neutrons the situation is somewhat different.
For $^{54}$Fe, little additional coherence is found beyond the
mean-field value, reflecting the closed-subshell neutron structure
of this nucleus. For the two other isotopes, the neutron pairing exhibits
two large eigenvalues. Although the larger one exceeds the mean-field
value and signals noticeable additional coherence across the subshells,
the existence of a second coherent eigenvalue shows the limits of the
BCS-like pairing picture.

In Fig. 3 we compare the SMMC pairing
strengths $P^J$ (Eq. (7))
 for the $^{54}$Fe ground state (calculated at
$T=0.5$ MeV) with the mean-field values. Although the largest values
for $P^J$ are found for pairs with larger $J$-values
(e.g. $J=2$ and 4 for isovector pairs), these values simply reflect
the larger combinatorial possibilities to make these $J$-pairs in our model
space, as the mean-field values are 
close to the SMMC values, i.e. there are no true correlations. 
Fig. 3 also exhibits the expected odd-even
staggering: in the isovector pairing channels, pairs with even $J$ 
are more likely than those with odd $J$, while it is vice versa for
isoscalar pairs. It is also obvious from Fig. 3 that the most
significant physical difference between the SMMC 
and mean-field pairing occurs in the isovector $J=0$ 
proton-proton and neutron-neutron channels.
Here, the SMMC calculations exhibit a significant excess of pairing correlations
over the mean field, reflecting the well known coherence in the ground states
of even-even nuclei. As the nucleus $^{54}$Fe is semimagic for
neutrons, the excess is larger for protons than for neutrons. Due to the
sum rule (8), which is approximately fulfilled for the $^{54}$Fe ground state,
the excess of $J=0$ pairs
is counterbalanced by correlation deficiencies in the other isovector pairs.

In the following we will discuss the physically important 
pairing correlation as defined in Eq. (10). Fig. 4 shows the
temperature dependence of the pair correlations for selected pairs, which,
as we will see in the next subsection, play an important role in
understanding the thermal behavior of the moment of inertia and the
total $M1$ and Gamow-Teller strengths.

The most interesting behavior is found in the $J=0$ proton and neutron pairs.
As mentioned above, the large excess of this pairing at low
temperatures reflects the ground state coherence of even-even nuclei.
With increasing temperature, this excess diminishes, vanishing at
around $T=1.2$ MeV. This behavior is in agreement with the 
pairing phase transition deduced
in Ref. \cite{temp} and
above from the temperature dependence of the BCS-like monopole pairs.
We observe further from Fig. 4 that the temperature dependence
of the $J=0$ proton-pair correlations is remarkably independent
of the nucleus, while the neutron pair correlations show interesting
differences. First, the correlation excess is smaller in the semimagic
nucleus $^{54}$Fe than in the others. Among the iron isotopes, the
neutron $J=0$ correlations vanish at higher temperatures
with increasing neutron number. More quantitatively,
we have fitted the correlation excess to a Fermi function,
\begin{equation}
P_{\rm corr}^J (T) = P_0 \left[ 1 + \exp {\left(-\frac{(T-T_0)}{\Delta T} 
\right) } \right]^{-1}.
\end{equation}
We then find (in MeV)
$T_0=0.82\pm0.13, \Delta T =0.13\pm0.07$ for $^{54}$Fe,  
$T_0=0.96\pm0.03, \Delta T =0.11\pm0.02$ for $^{56}$Fe, and  
$T_0=1.07\pm0.04, \Delta T =0.18\pm0.04$ for $^{58}$Fe, while 
the fit for $^{56}$Cr
is very similar to that for $^{58}$Fe,  
$T_0=1.06\pm0.04, \Delta T =0.14\pm0.03$.
Similar fits to the $J=0$ proton pairing excess results in  
$T_0\approx0.9\pm0.04, \Delta T =0.12\pm0.03$ for all nuclei.
The vanishing of the $J=0$ proton and neutron pair correlations is
accompanied by a significant increase in the correlations of the other pairs.

The isovector $J=0$ proton-neutron correlations are positive at all temperatures
for all four nuclei, with a slight excess at low
temperatures that increases by about a factor 3 after the $J=0$
proton and neutron pairs have vanished. The correlation peak is reached
at higher temperatures with increasing neutron number 
(about 1.3 MeV
for $^{54}$Fe, 1.6 MeV in $^{58}$Fe), while the
peak height decreases with neutron excess.

The isoscalar proton-neutron $J=1$ pairs show an interesting temperature
dependence. At low temperatures, when the nucleus is still dominated
by the $J=0$ proton and neutron pairs, the isoscalar proton-neutron 
correlations show a
noticeable excess. But more interestingly, they are roughly constant and
do not directly reflect the vanishing of the $J=0$ proton and neutron pairs.
However, at $T>1$ MeV, when the proton and neutron-pairs are broken, 
the isoscalar
$J=1$ pair correlations significantly increase and have their maximum
at around 2 MeV, with peak values of about twice the correlation excess
in the ground state. A similar (but milder) increase is observed in the
other isoscalar correlations.
In contrast to
the isovector $J=0$ proton-neutron pairs, the correlation peaks
are reached at lower temperatures with increasing neutron excess. We also 
observe that these correlations die out rather slowly with temperature.

In summary, our SMMC calculations 
of nuclei around $A=56$
support the following temperature
hierarchy of pairing correlations. 
In the ground state and at low temperatures, $J=0$ proton and neutron
pair correlations dominate. When these vanish at temperatures around
1 MeV, proton-neutron correlations become more important, with the
isovector proton-neutron correlations, in our model space, vanishing at
temperatures around 2 MeV. At high temperatures, isoscalar proton-neutron
correlations dominate.

\subsection{Pairing correlations and observables}

In this subsection we will discuss 
the relationship between pairing correlations and
the temperature response of the $M1$, and Gamow-Teller strengths,
as well as of the moment of inertia. 
As in our definition of pairing, we will focus on the behavior of these
observables beyond the mean-field values. In terms of the 
neutron and proton occupation numbers
$n_n$ and $n_p$, respectively,
the mean-field values for these observables are given by
\begin{equation}  
\langle I \rangle_{\rm mf} 
= \frac{3}{T} \sum_{i} \langle i|J^2|i \rangle 
\left[ 
n_{p}(i) {\bar n_p} (i) +
n_{n}(i) {\bar n_n} (i) \right]
\end{equation}
\begin{equation}
B(M1)_{\rm mf} = \sum_{i,j} 
\left[ | \langle i|{\vec \mu}(p)|j \rangle|^2 
n_{p}(j) {\bar n_p} (i) +
| \langle i|{\vec \mu}(n)|j \rangle |^2 
n_{n}(j) {\bar n_n} (i) \right]
\end{equation}
\begin{equation}
B(GT_+)_{\rm mf} = \sum_{i,j} 
| \langle i|{\vec \sigma} \tau_+|j \rangle |^2 
n_{p}(j) {\bar n_n} (i)
\end{equation}  
where the blocking factors are
\begin{equation}
{\bar n_p} (i) = 1 - \frac{n_p(i)}{2j_i+1};
\end{equation}
\begin{equation}
{\bar n_n} (i) = 1 - \frac{n_n(i)}{2j_i+1};
\end{equation}
and
${\vec \mu} (p)$
and ${\vec \mu} (n)$ are the proton and neutron 
magnetic moments.

The SMMC results for all three observables are significantly smaller than
the mean-field values. To quantify this suppression, 
we introduce  quenching
strengths 
\begin{eqnarray}
\langle I \rangle_{\rm quench} 
&=& \langle I \rangle_{\rm mf} 
- \langle I \rangle; \\
B(M1)_{\rm quench} &=& B(M1)_{\rm mf} - B(M1) ; \\ 
B(GT_+)_{\rm quench} &=& B(GT_+)_{\rm mf} - B(GT_+).
\end{eqnarray}

We expect the quenching strength of the observables to be larger when the
relevant pairing correlations are. Fig. 5 shows the
temperature dependence of the quenching strengths for the three observables
and relates it to selected pairing correlations, where the latter
are given in arbitrary units.
The following observations are evident and apply to all four nuclei:

The quenching of the moment of inertia shows a sharp drop at around
$T=1$ MeV and then slowly diminishes; the SMMC and mean-field values
are identical for $T>2$ MeV.
The drop at $T\approx1$ MeV clearly signals a relation to the
pairing phase transition observed in the $J=0$ proton and neutron
correlations (see Fig. 4). From the definition 
$I \sim \langle J^2 \rangle = \langle (J_n+J_p)^2 \rangle$, where
$J_p, J_n$ are the proton and neutron angular momenta, respectively,
we expect that the moment of inertia is sensitive to the proton,
neutron and proton-neutron correlations. This expectation is confirmed
in Fig. 5, which shows that the temperature dependence of the quenching
is nearly identical to the sum of the isovector $J=0$
pair correlations. The steep drop at $T\approx 1$ MeV is thus due
to the $J=0$ proton and neutron pair correlations. We note that the
moment of inertia unquenches more slowly with temperature than the
proton and neutron $J=0$ pairs. This difference is due to the isovector
$J=0$ proton-neutron correlations.

The quenching of the M1 strength decreases significantly near $T=1$ MeV. 
Following a minimum at $T\approx1.3$ MeV, the quenching increases again
and then, after a maximum near 2 MeV, decreases slowly. While the drop
at $T=1$ MeV again signals an association with the 
$J=0$ proton and neutron pairing phase transition, the maximum at 2 MeV 
as well as the slow dying out clearly resembles the temperature dependence
of the isoscalar $J=1$ proton-neutron correlations. These two correlations
reflect the two components in the M1 strength. The orbital part is sensitive
to the $J=0$ proton pairing correlations (the gyromagnetic moment $g_l$
is zero for neutrons), while the quenching of the spin component 
is dominated by
isoscalar proton-neutron correlations \cite{Engel,GT}. This conjecture
is in fact nicely confirmed, as for all four nuclei the temperature
dependence of the quenching of the M1 strength is well described by 
that of the sum of the $J=0$ proton and isoscalar $J=1$ proton-neutron
correlations. 

Our calculation reveals a close (linear) relation between the orbital part
of the M1 strength and the moment of inertia at low temperatures, where
both quantities are dominated by isovector $J=0$ pairing correlations.
Such a linear relation is experimentally known for heavier nuclei
and is supported by QRPA calculations \cite{Richter1,Coster,Iudice}.

The quenching of the $B(GT_+)$ strength is roughly constant at low
temperatures ($T<1$ MeV). It then increases slightly, before it slowly
dies out following a maximum at $T \approx 2.5$ MeV. 
Fig. 5 shows that 
the temperature dependence of the
Gamow-Teller strength is driven by the isoscalar proton-neutron
correlations, 
as the rough
constancy at low temperatures, the maximum at around $T=2.5$ MeV
and the slow decrease of the correlations towards higher temperatures
of the $B(GT_+)$ quenching qualitatively mimics the temperature
dependence of the isoscalar $J=1$ proton-neutron correlations.
With increasing neutron excess the maximum in the $B(GT_+)$ quenching
at $T \approx 2.5$ MeV is weakened, in qualitative agreement with the
$J=1$ proton-neutron correlations. 
 
\section{Summary}

In a previous SMMC study 
of the nucleus $^{54}$Fe, a sharp rise in the moment of inertia and the
M1 strength at a temperature $T \approx 1$ MeV were related to
the vanishing of the BCS-like proton and neutron Cooper pairs.
In this paper we have investigated in more details
this phase transition and its connection
to the temperature dependence of the moment of inertia
and of the M1 and Gamow-Teller strength. To do so we
have performed SMMC calculations
for the iron isotopes
$^{54,56,58}$Fe and $^{56}$Cr within in the complete
$pf$-shell using the KB3 interaction. To study the pairing correlations
within the shell model, we have defined a generalized pair matrix for
isovector and isoscalar pairs for a given angular momentum and have
proposed the sum of the eigenvalues of this matrix (trace) as a measure of the
pairing strength. Using the same formalism, but
replacing the two-body pair operators in the
generalized pair matrix by occupation numbers we have derived the
mean-field pairing strength. The physically relevant pair correlations
are then defined as the difference of the SMMC and mean-field pairing
strengths.

A detailed investigation of the correlations in the various
pairing channels gives the following qualitative picture. 

At low temperatures the SMMC calculations show a strong excess of
$J=0$ proton and neutron pairs
over the mean-field
values. This excess drops sharply at around $T=1$ MeV for all four nuclei
studied here, confirming the pairing phase transition conjectured
in Ref. \cite{temp}. After the vanishing of the dominating $J=0$ pairs,
proton-neutron correlations show an excess over the mean-field values.
We find that isoscalar proton-neutron correlations persist to higher
temperatures than the isovector correlations and, within our model space,
vanish slowly at temperatures larger than 3 MeV. We observe some evidence
that the pairing correlations depend on the neutron excess of the nucleus.
This nuclear isospin-dependence warrants more detailed explorations.

The temperature hierarchy of the pairing correlations and the $J=0$
proton and neutron pairing phase transition manifests itself in the
temperature dependence of the three observables we studied in detail.
The moment of inertia of the nucleus correlates with the sum of the
isovector $J=0$ pair correlations. In particular, we find that  the
pairing phase transition is connected with a sharp rise of the moment
of inertia. Due to its orbital and spin components,
the M1 strength unquenches generally
in two steps: the orbital part unquenches at the $J=0$ proton and neutron
pairing phase transition, while the spin part slowly unquenches at higher
temperatures, in concert 
with the decrease of the isoscalar proton-neutron
correlations.
The temperature dependence of the Gamow-Teller quenching $B(GT_+)$ is well
approximated by that of the isoscalar $J=1$ proton-neutron pairs.

Although the present calculations have been performed in model spaces
larger than any imagined tractable until only recently, 
the quantitative  results
suffer from finite model space limitations at temperatures greater about
1.5 MeV. Although we believe that the  qualitative results presented
in this paper will persist in calculations performed in even larger model
spaces, such work is
certainly warranted for a quantitatively reliable
description of the unquenching of the isoscalar correlations at high
temperatures. Such calculations are feasible within the SMMC approach and,
as a next step, we are planning studies in the complete
($pf$+$sdg$) shells. 

\acknowledgements

This work was supported in part by the National Science Foundation,  
Grants No. PHY94-12818 and PHY94-20470.
Oak Ridge National Laboratory is managed by Lockheed Martin Energy
Research Corp. for the U.S. Department of Energy under contract number
DE-AC05-96OR22464.
We are grateful to P.~Vogel for helpful discussions. Computational
cycles were provided by
the Concurrent Supercomputing Consortium and by the VPP500,
a vector parallel processor at  
the RIKEN supercomputing facility; we thank Drs. I. Tanihata and
S. Ohta for their assistance with the latter.

\begin{figure}
\caption{ Temperature dependence of the moment of inertia (left top),
the $B(M1)$ strength (left bottom), the $B(GT_+)$ strength (right top),
and the expectation values of the BCS proton and neutron
pairing strength (right bottom). SMMC results are shown for
$^{54}$Fe (part a), $^{56}$Fe (part b), $^{58}$Fe (part c), and
$^{56}$Cr (part d).}
\label{fig1}
\end{figure}

\begin{figure}
\caption{Diagonal elements (upper) and eigenvalues (middle)
of the pair matrix $M$
in the three isovector $J=0^+$ and the isoscalar $J=1^+$ channels,
as calculated for $^{54,56,58}$Fe at $T=0.5$ MeV.
Eigenvalues of the pair matrix on the mean-field level (see text) 
are shown in the lowest panels.}
\label{fig3}
\end{figure}

\begin{figure}
\caption{ Comparison of the SMMC ($P^J$) and mean-field
($P^J_{\rm mf}$) pairing strengths in the $^{54}$Fe ground state
for selected pairing channels.}
\label{fig2}
\end{figure}

\begin{figure}
\caption{ Temperature dependence of the pair correlations $P^J_{\rm corr}$
as defined in Eq. (10) for selected pairing channels.}
\label{fig4}
\end{figure}

\begin{figure}
\caption{Comparison of the temperature dependence
of the quenching of the physical observables (see Eqs. (18-20))
with related pairing
correlations for $^{54,56}$Fe (part a) and $^{58}$Fe, $^{56}$Cr (part b). 
Upper: moment of inertia and the sum of the isovector $J=0$ pairing
correlations, 
Middle: the M1 strength and the sum (solid line) 
of $J=0$ proton (dashed) and isoscalar $J=1$ proton-neutron (dotted)
correlations;
Lower: the $B(GT_+)$ strength and the isoscalar $J=1$ proton-neutron
correlations.
The pairing correlations have been scaled arbitrarily.}
\label{fig5}
\end{figure}

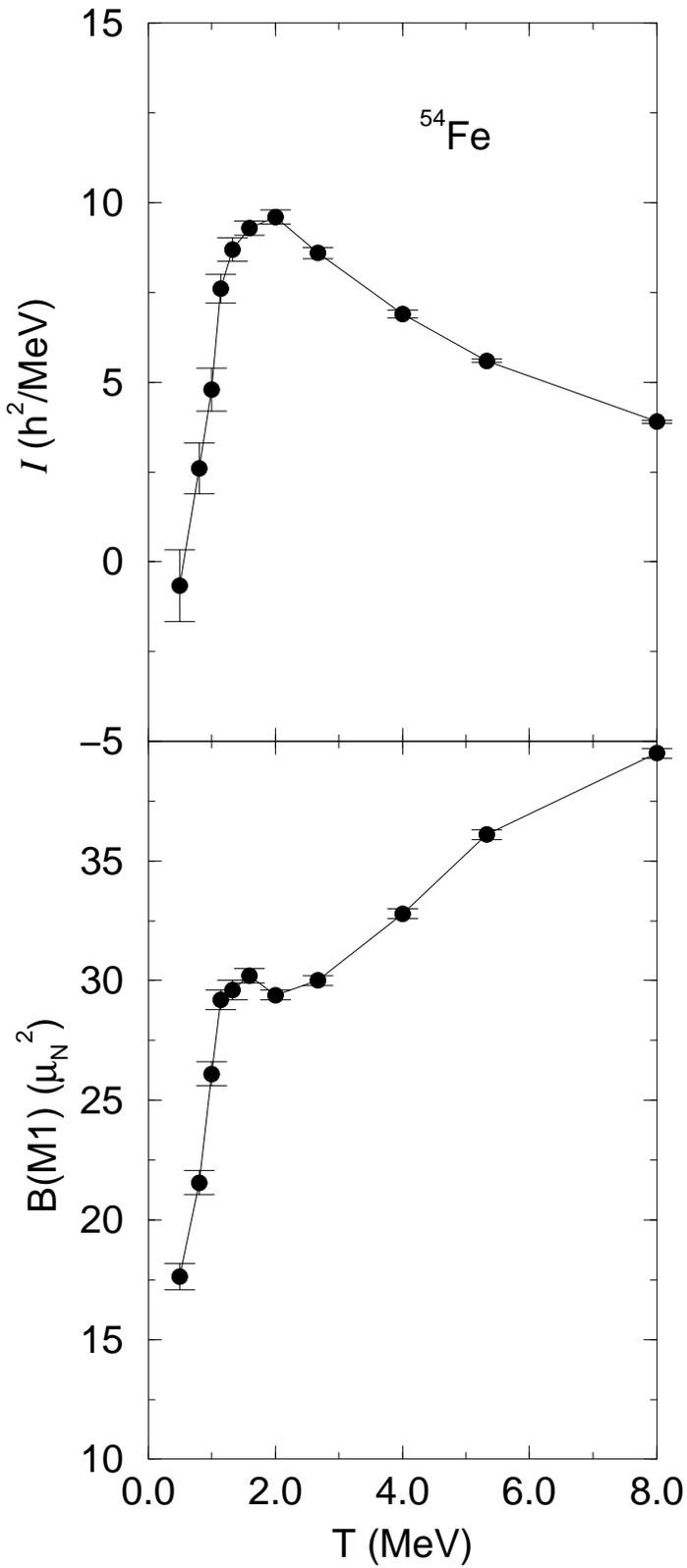
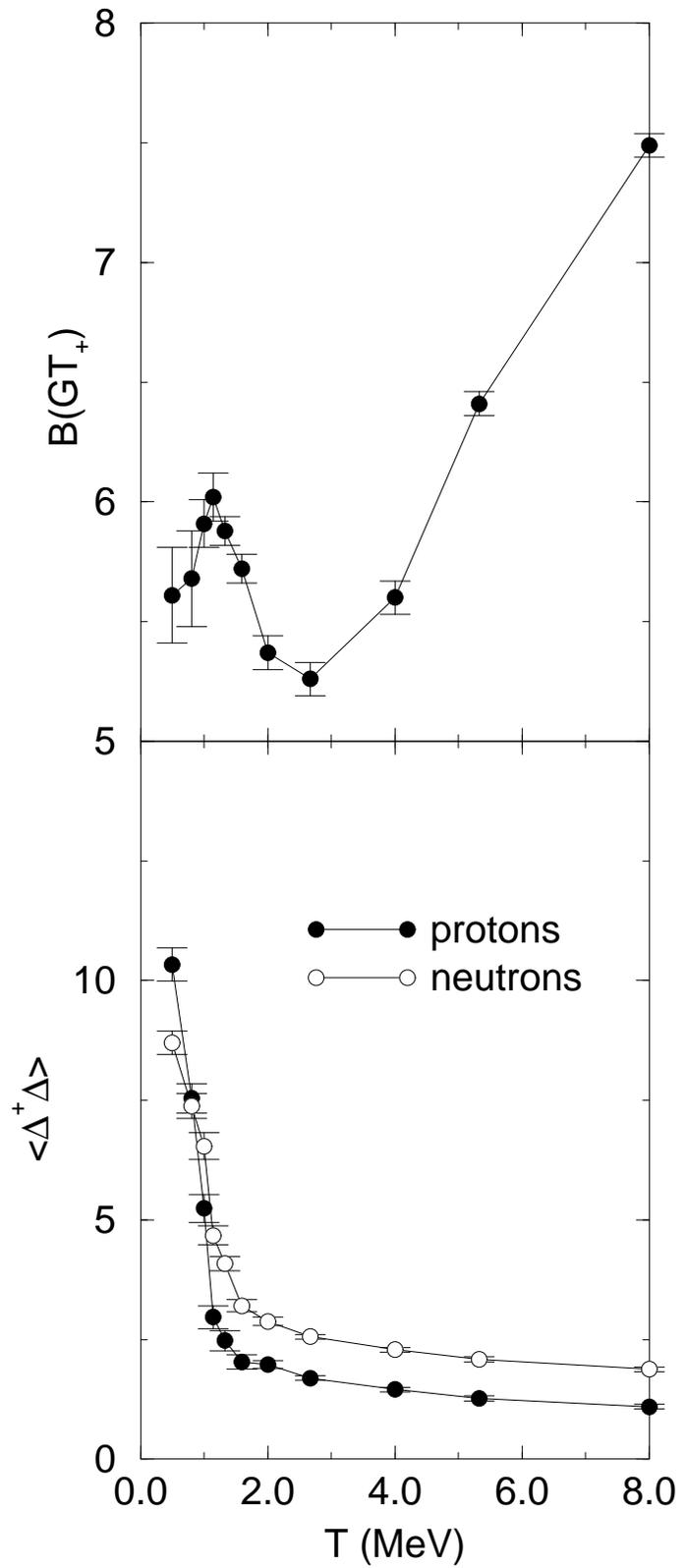
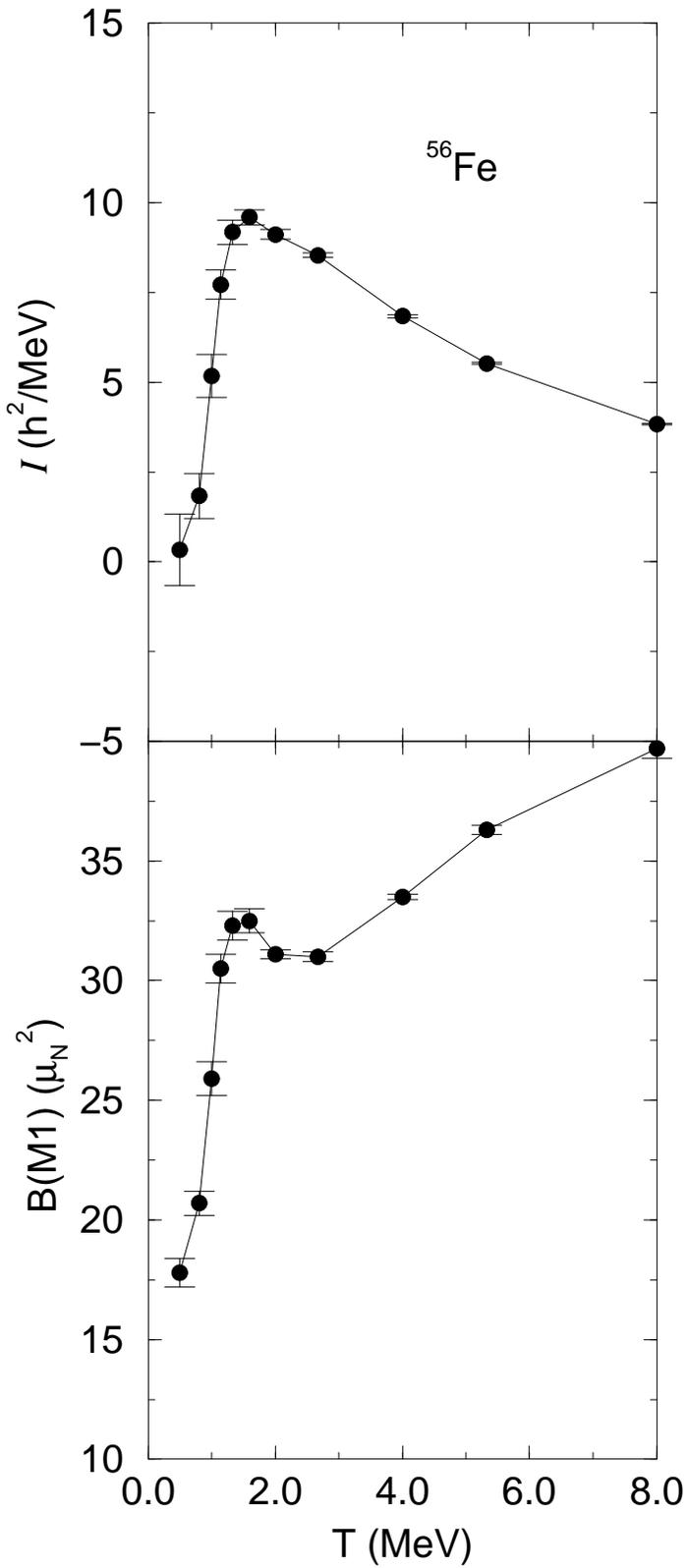
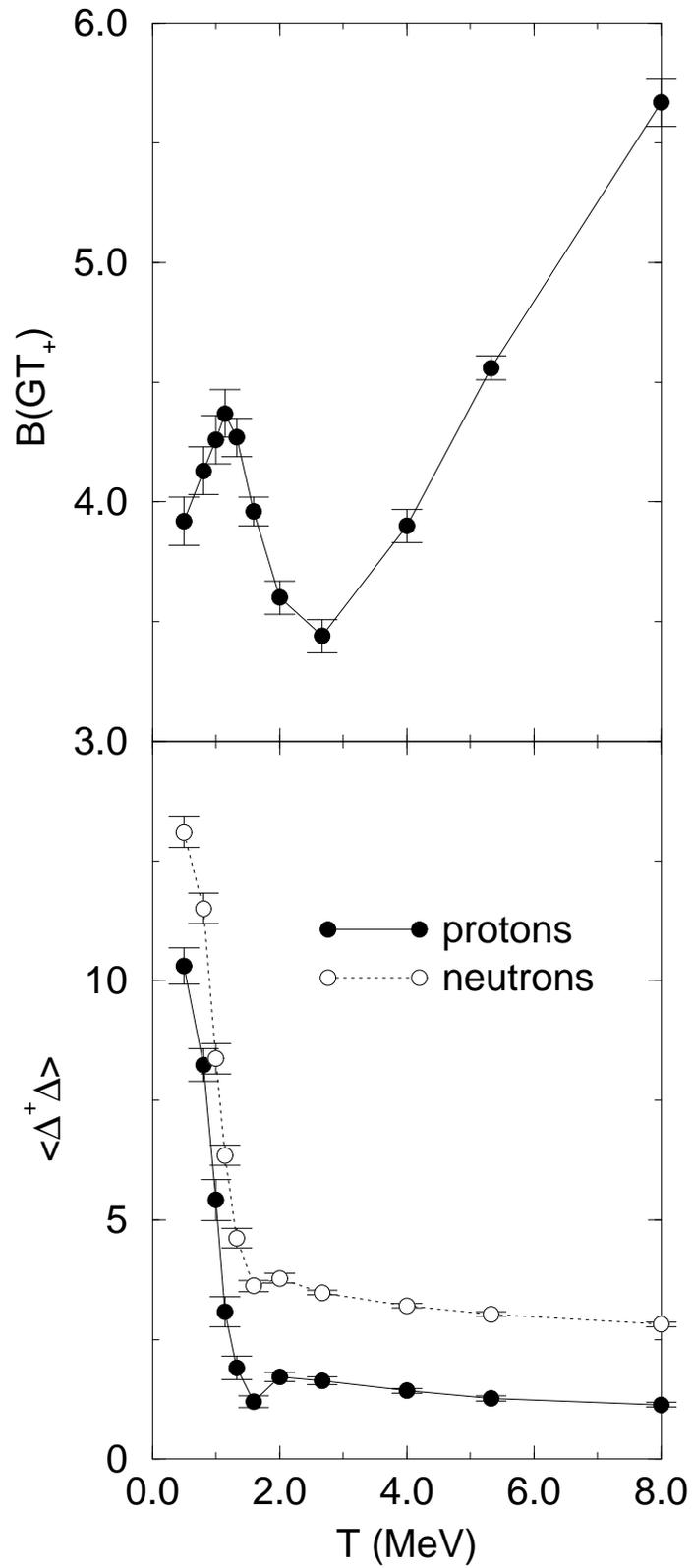
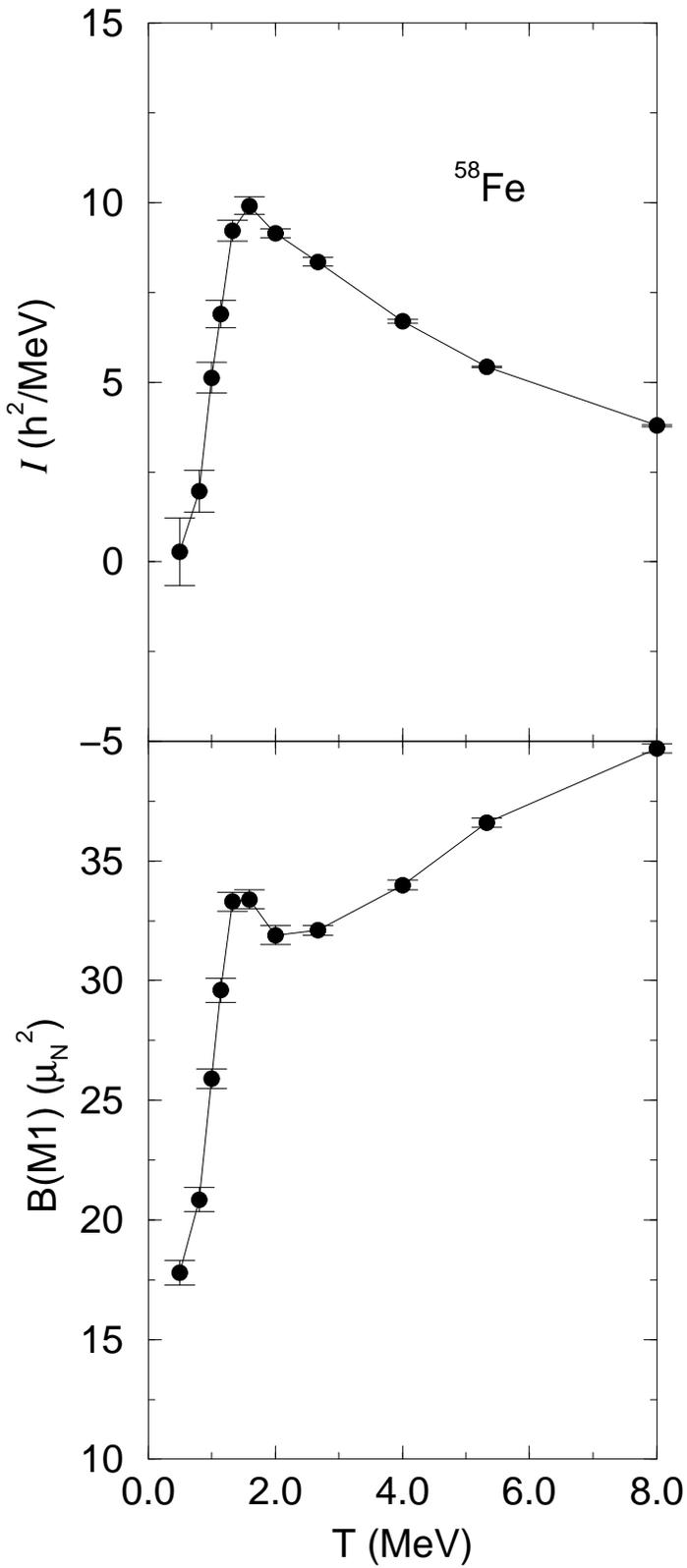
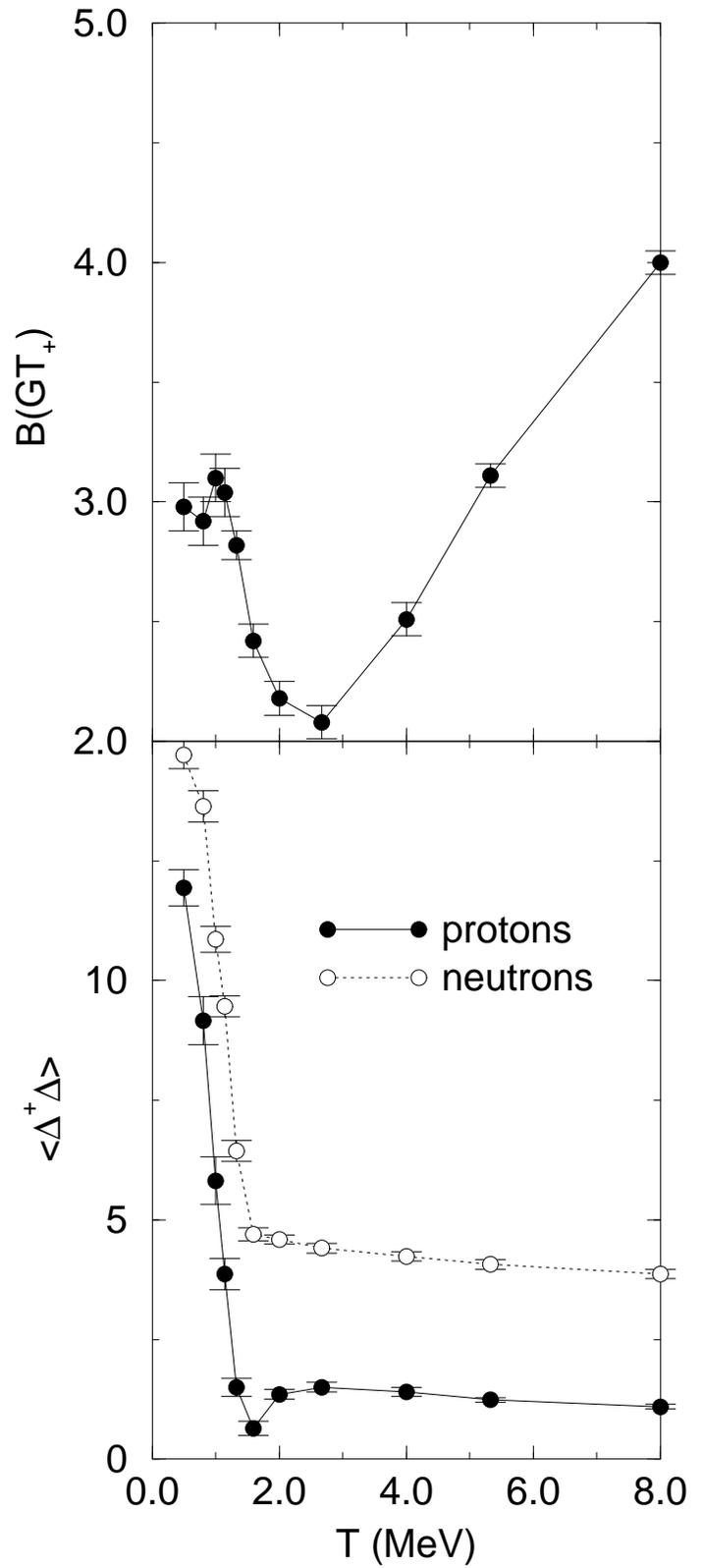
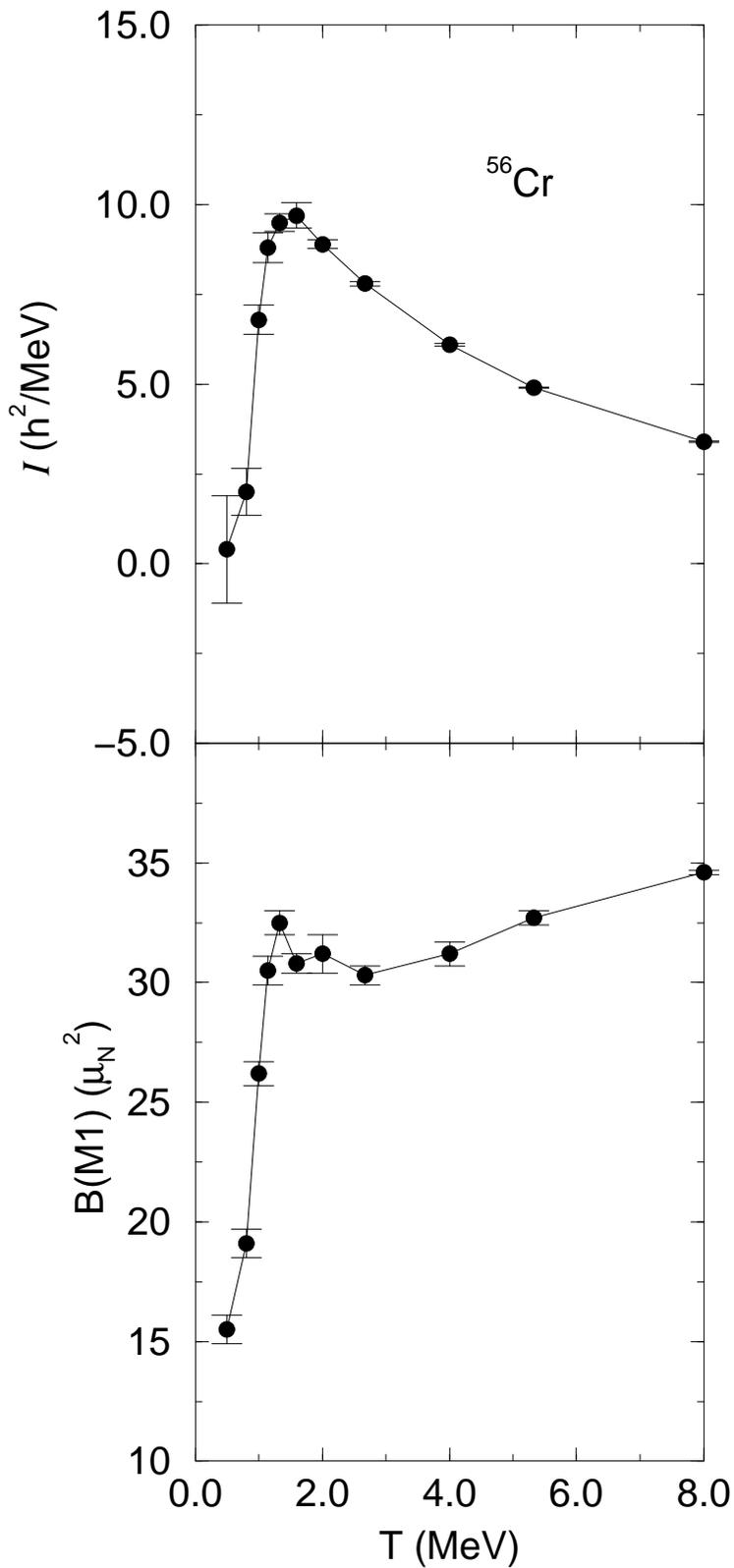
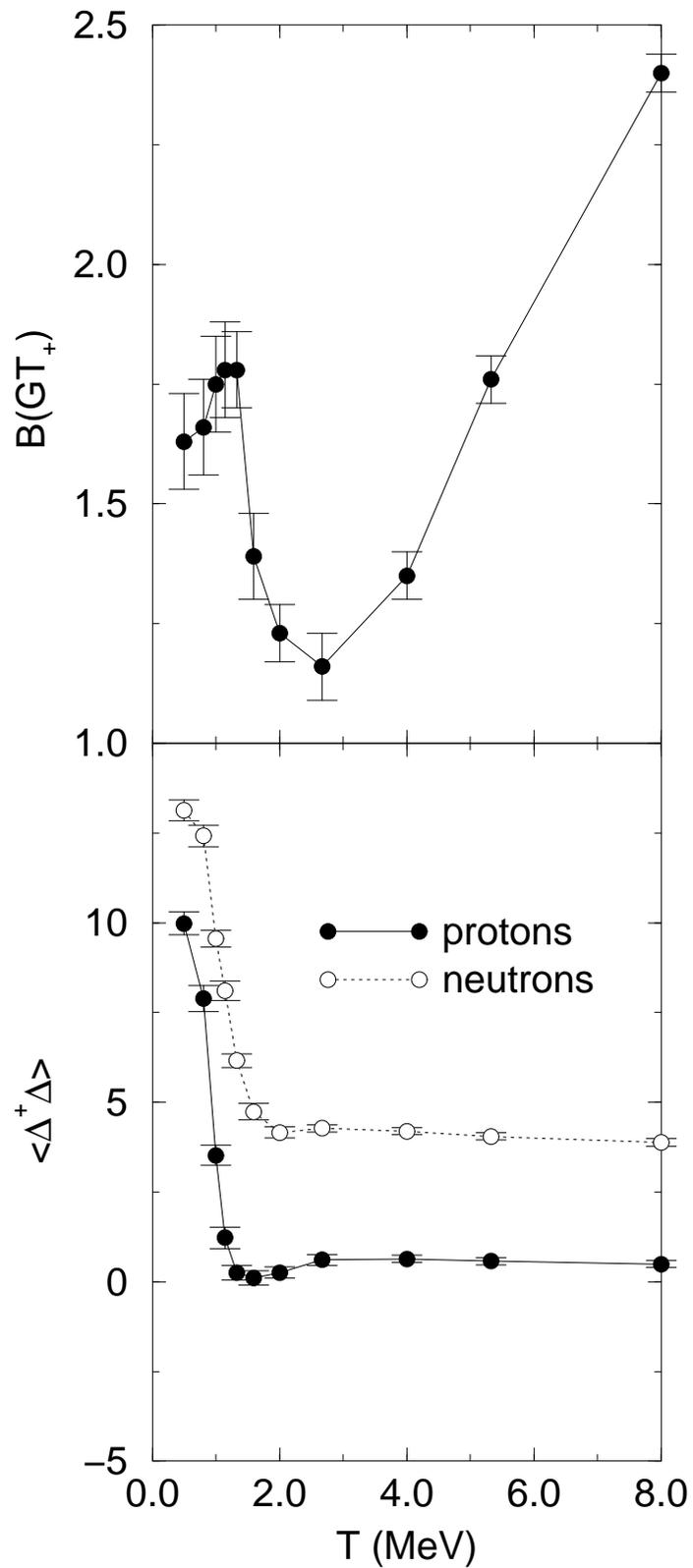
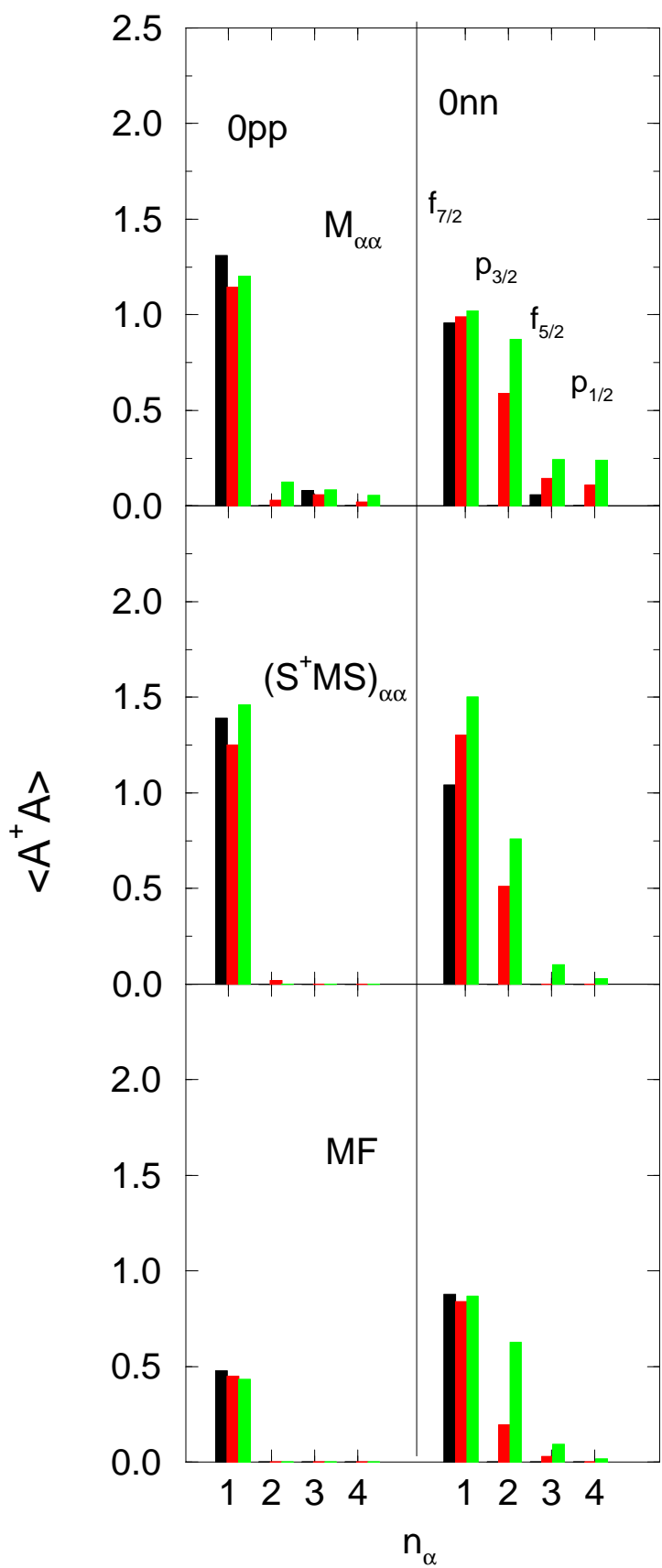
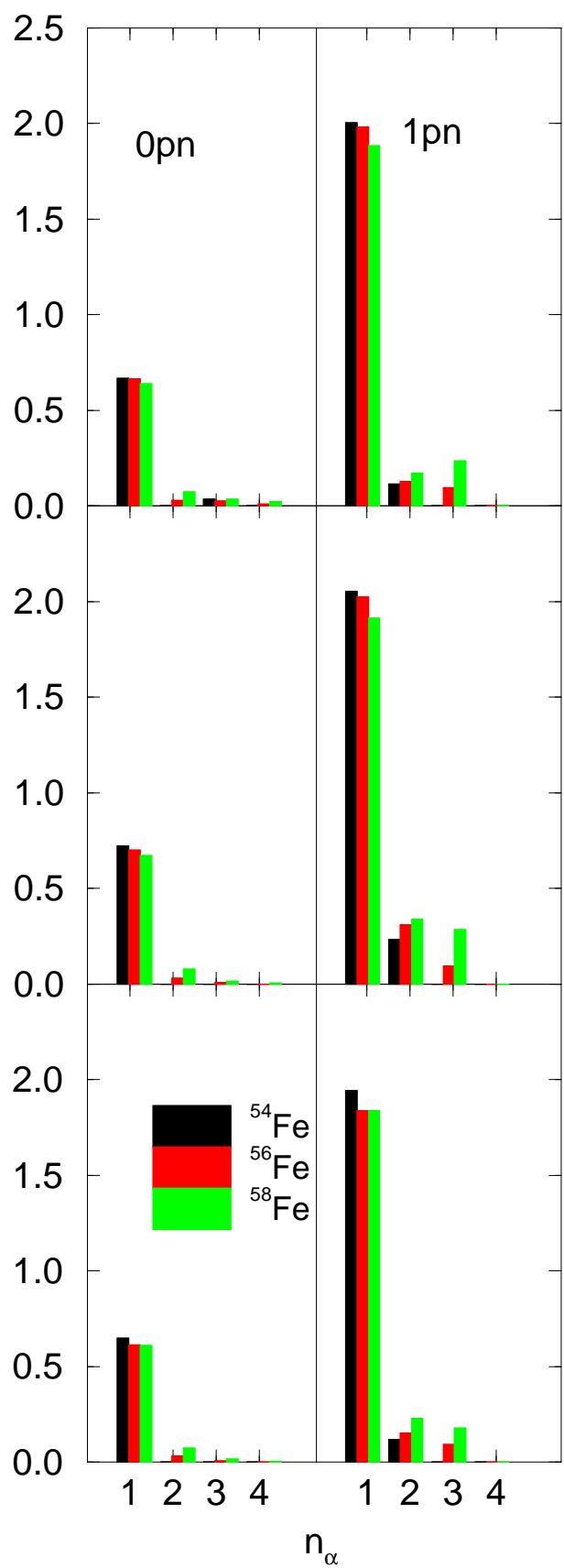
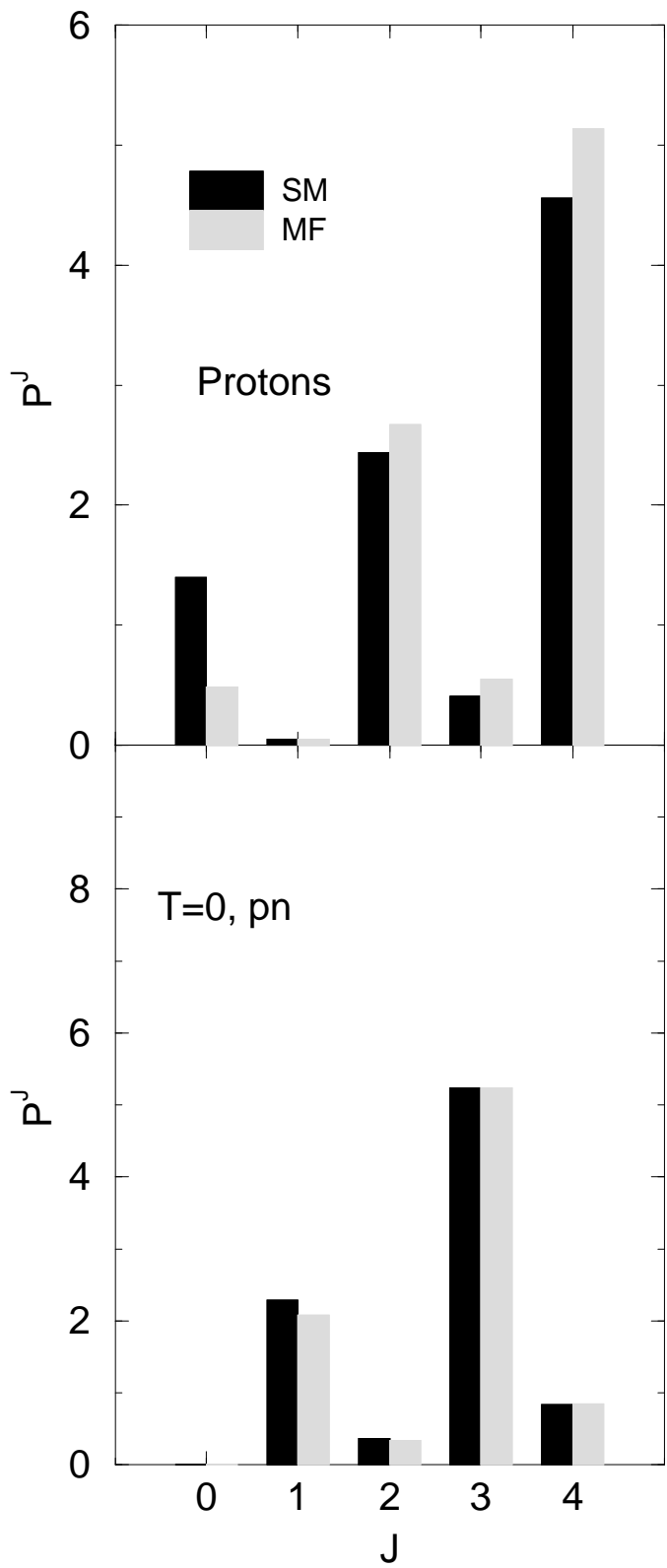
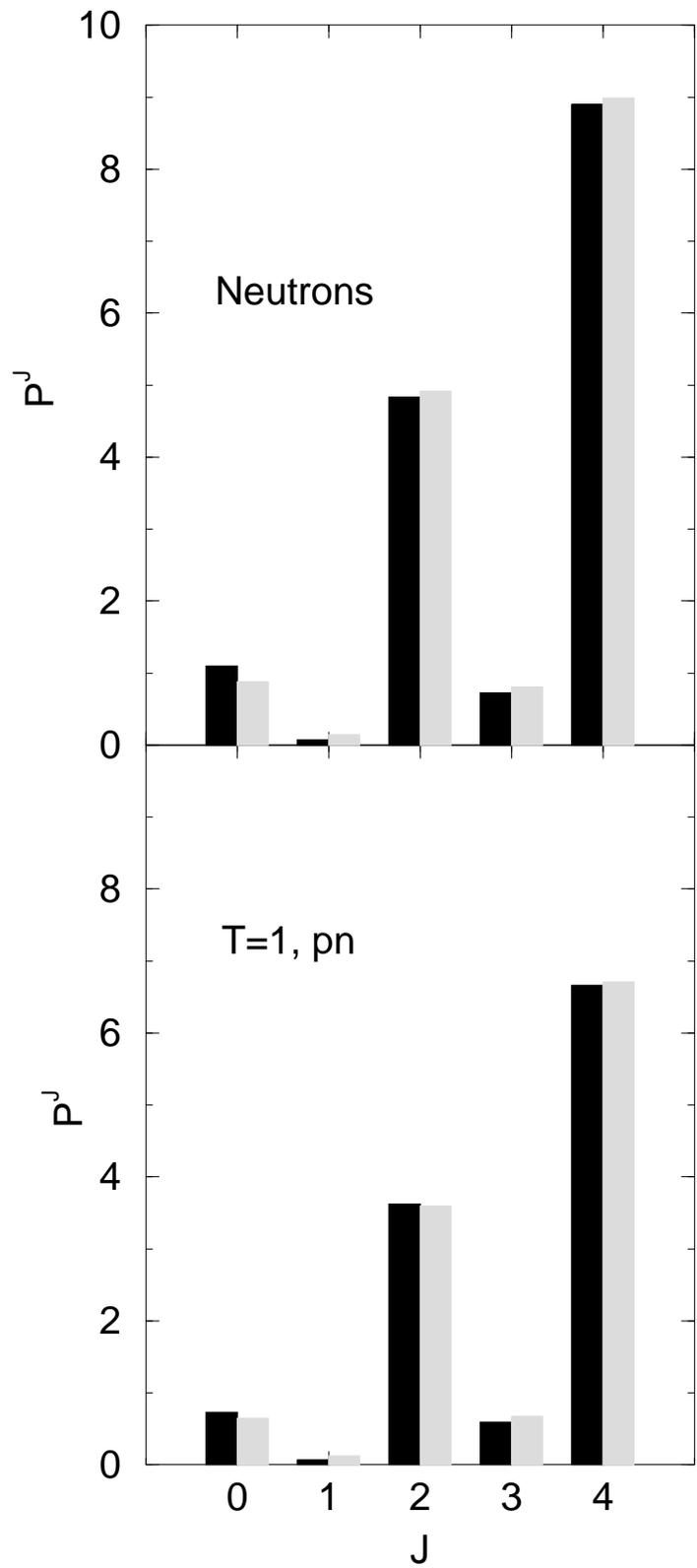
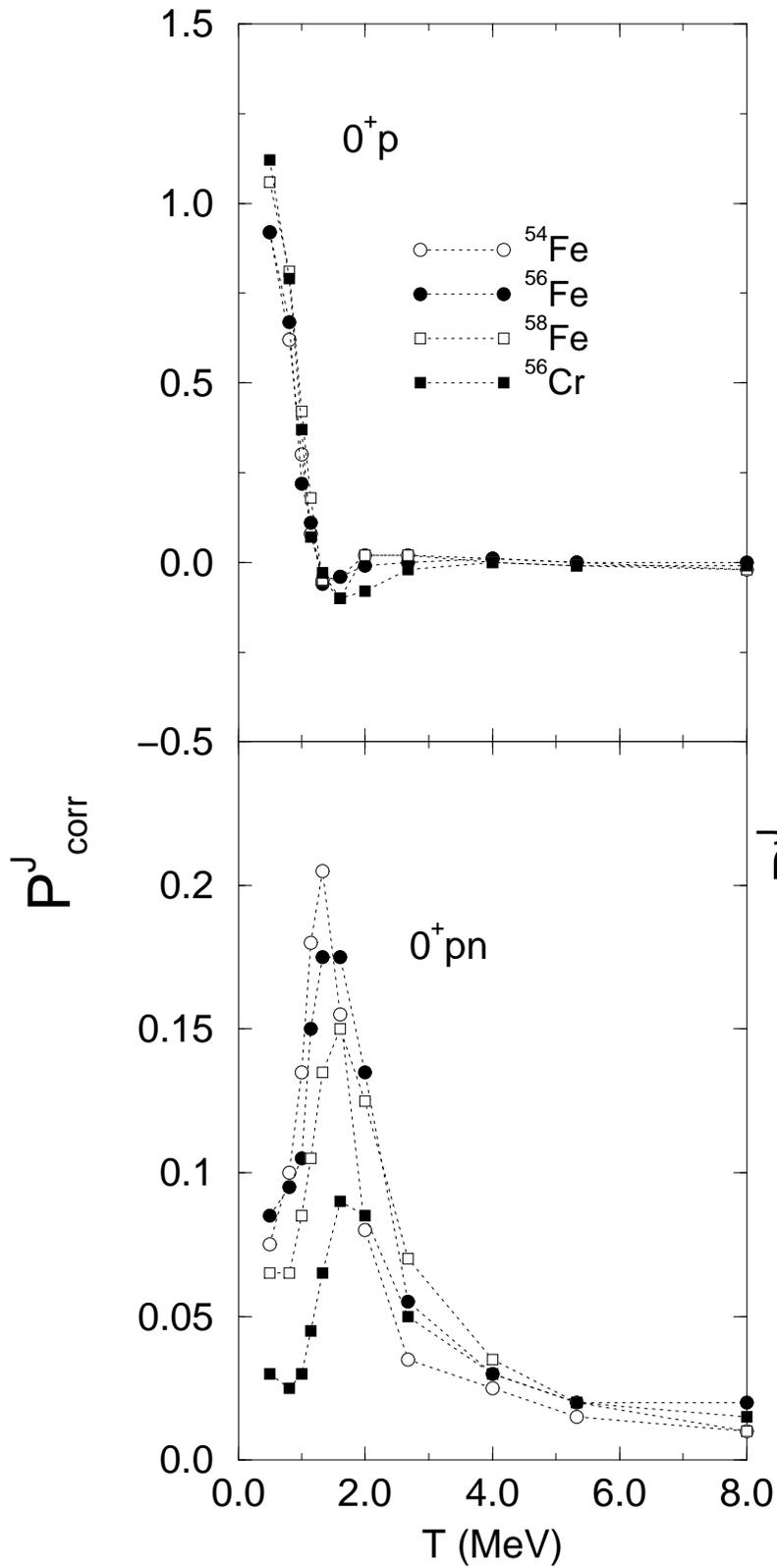
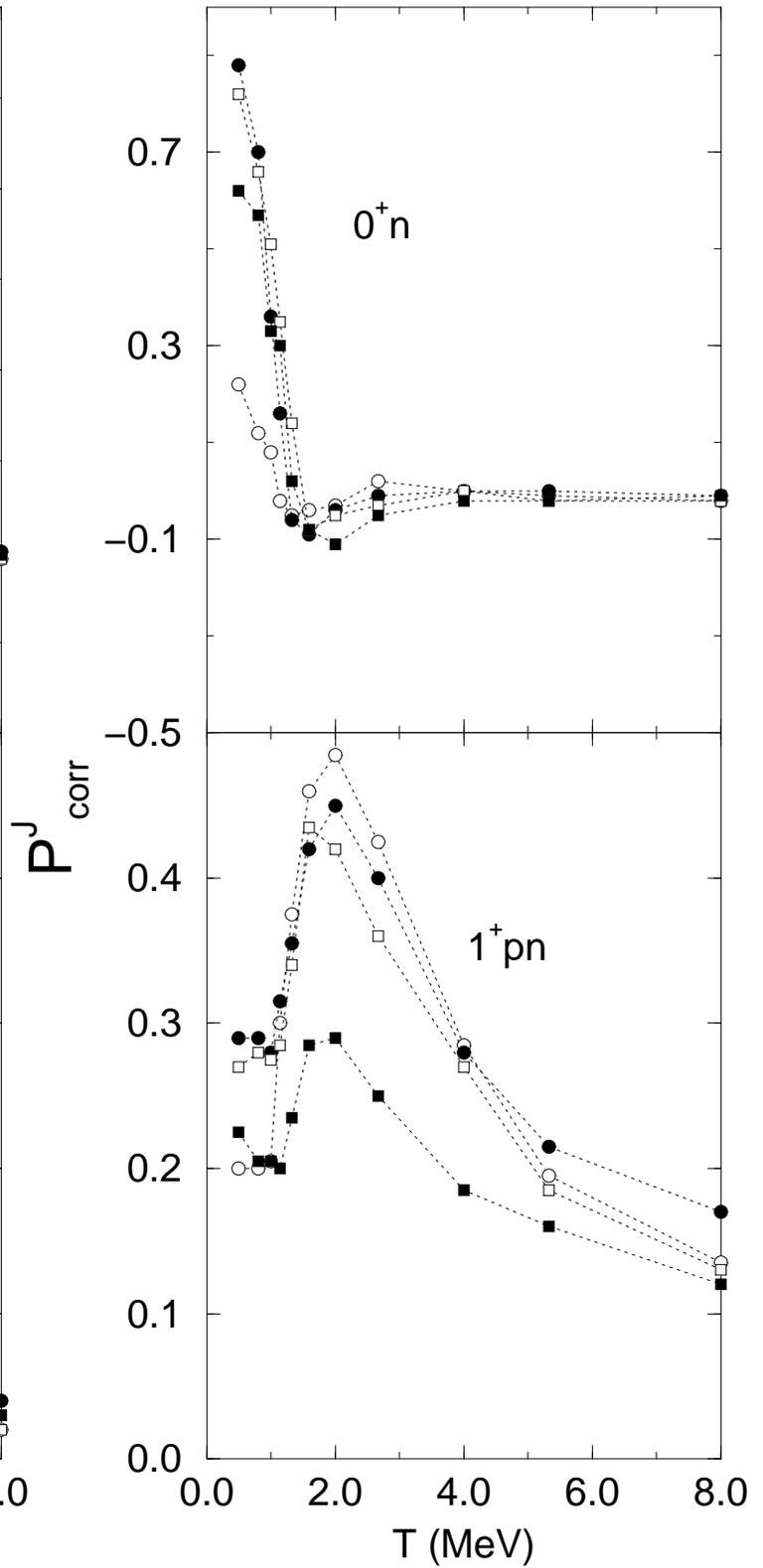
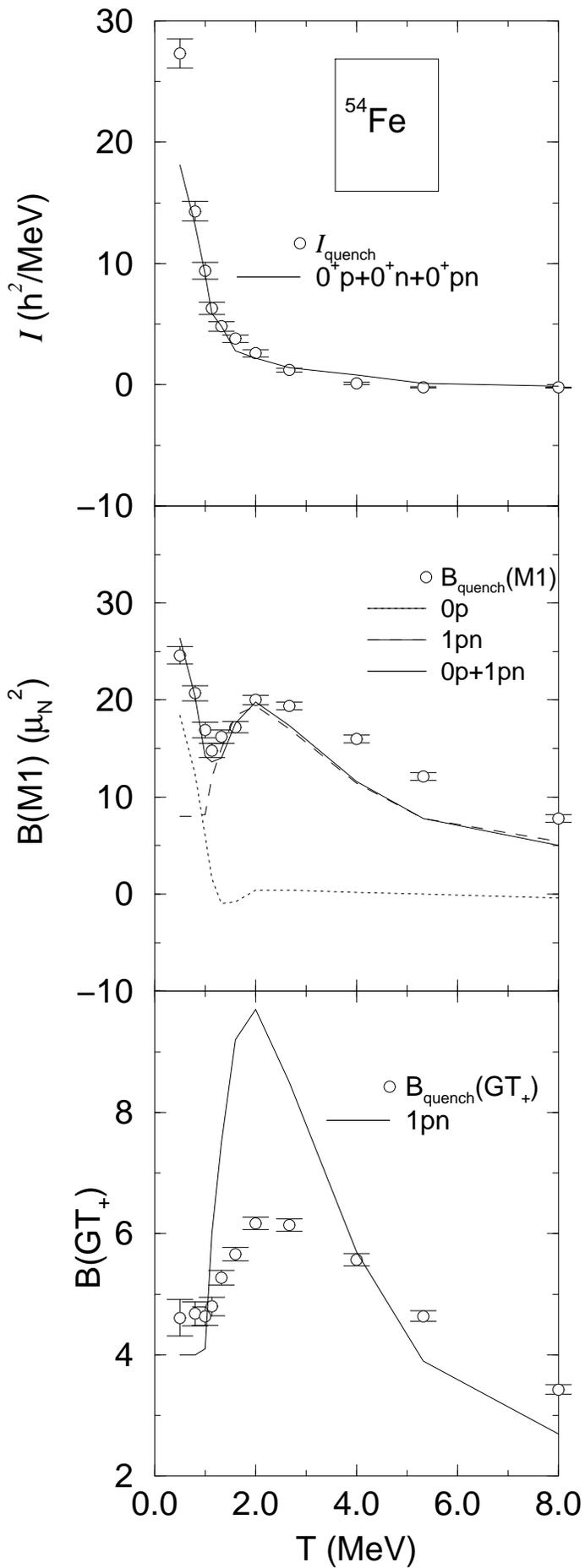
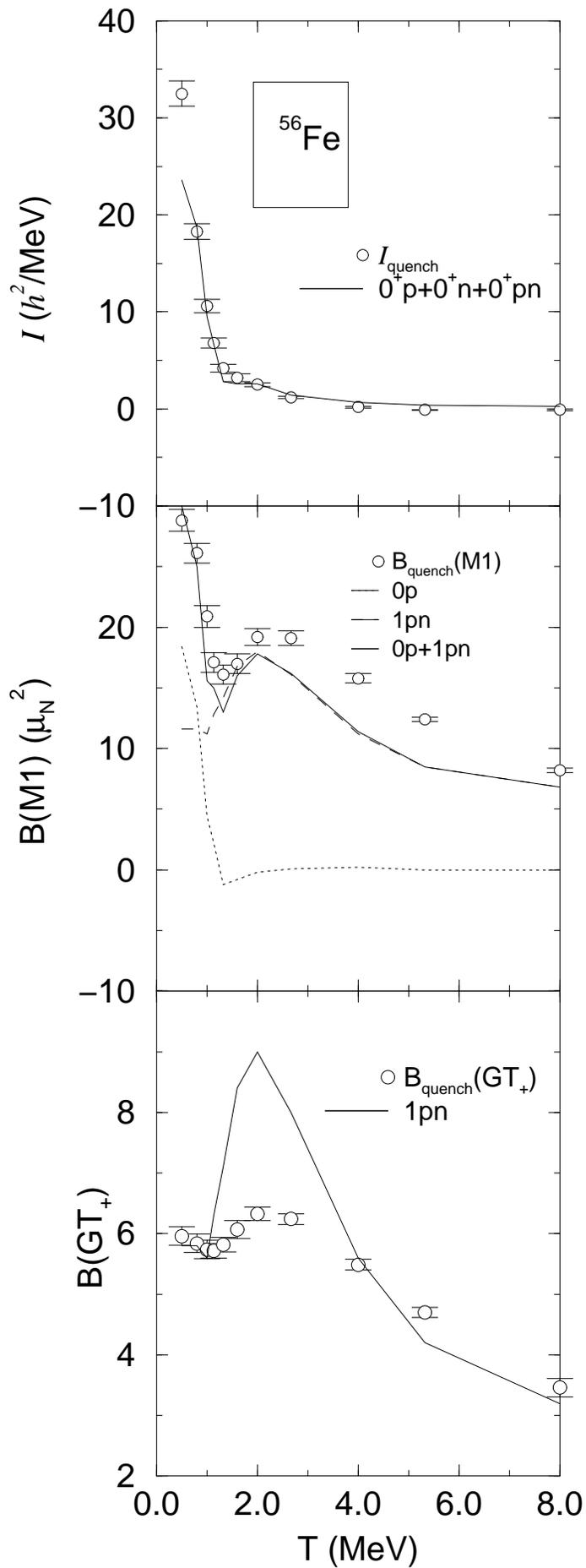
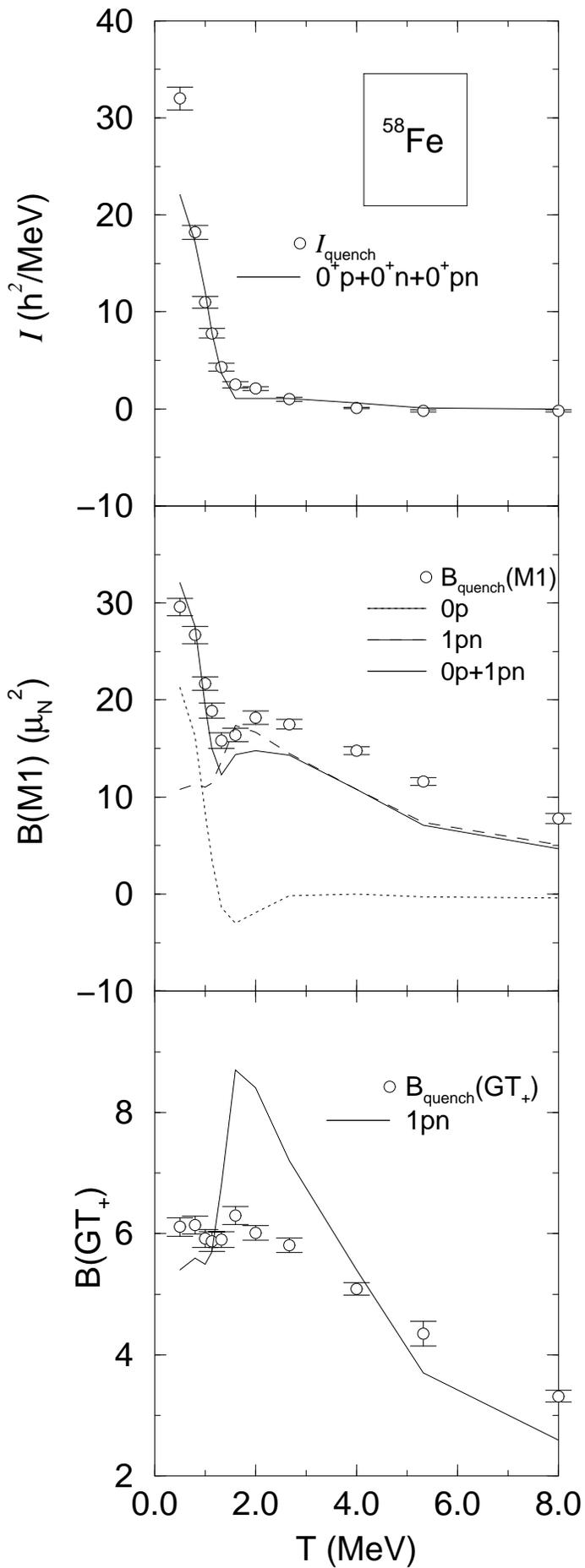
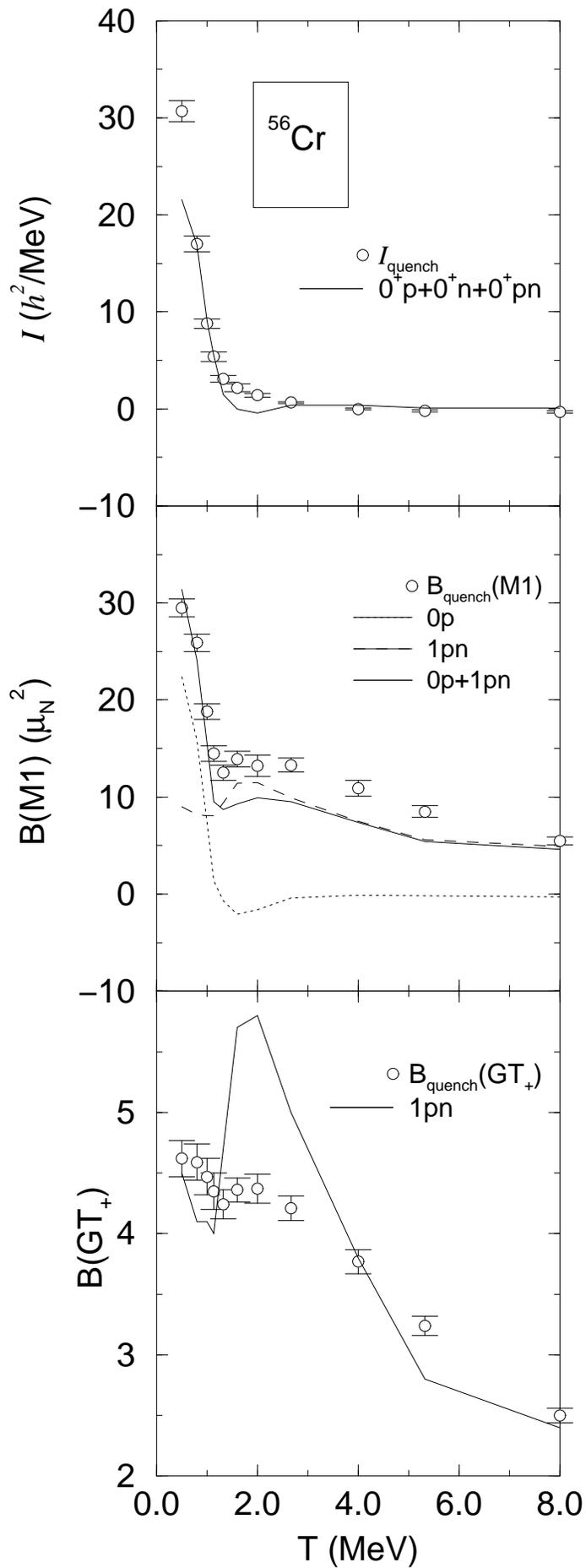
\end{document}